\documentclass{iau}
\usepackage{graphicx} 

\title[IAUS291.~~Glitches in southern pulsars] 
{Glitches detected in southern radio pulsars} 

\author[M. Yu] 
{Meng Yu$^{1, 2, 3}$
\thanks{Present address: 20A Datun Road, Chaoyang District, Beijing
  100012, P. R. China}}

\affiliation{$^1$School of Physics and State Key Laboratory of Nuclear
  Physics and Technology, Peking University \\
[\affilskip] $^2$CSIRO Astronomy and
  Space Science \\ 
[\affilskip] $^3$National Astronomical Observatories of China
  \\ [\affilskip] email: {\tt vela.yumeng@gmail.com}} 

\pubyear{2012}
\volume{291}  
\jname{\mbox{Neutron Stars and Pulsars: Challenges and Opportunities
    after 80 years}} \editors{J. van Leeuwen, ed.}
\begin{document}

\maketitle

\begin{abstract}
Parkes pulse arrival-time data for 165 radio pulsars spanning from
1990 to 2011 have been searched for period glitches. Forty-six events
out of the detected 107 glitches were found to be new
contributions to the entire glitch population which currently contains
approximately 400 events.
\keywords{stars: neutron - pulsars: general}
\end{abstract}


\firstsection 

\section{Introduction}\label{sect:intro}

Two types of timing irregularities have been hampering a full
interpretation of the rotation of neutron stars: timing noise -- a
random process seen in phase residuals and glitches -- a
discontinuous increase in the pulse frequency $\nu$. Glitches are
transient events; the most stringent upper-bound known for the
timescale of the rising edge of $\nu$, 40\,s, was obtained by
\cite[Dodson, McCulloch \& Lewis (2002)]{dml02} in a high
time-resolution observing program for the Vela pulsar. The frequency
jump caused by a glitch is actually small with the detected maximum
relative size of $\Delta\nu_{\rm g}/\nu\sim10^{-5}$~\cite[(Yuan et
  al. 2010; Manchester \& Hobbs 2011)]{ymw+10,mh11}. After a glitch,
the slow-down rate $|\dot\nu|$ often partially recovers exponentially,
followed by a long-term linear decrease; permanent changes in $\nu$
and its time derivatives are usually left. These widely observed
features are used to model glitches.

\section{Observations and Results}\label{sect:obs_res}

\begin{figure}
\begin{center}
\includegraphics[width=7cm,angle=-90]{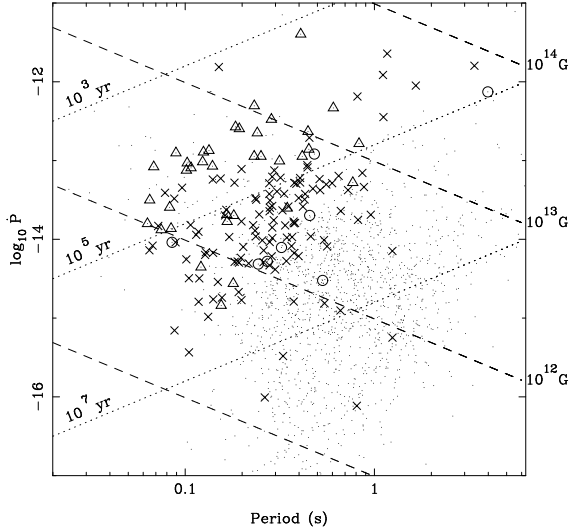}
\caption{$P$ -- $\dot{P}$ diagram showing the 165 pulsars in the
  sample. The various symbols indicate the pulsars where no glitch has
  been reported ($\times$), glitches were detected within our data
  span ($\triangle$) and glitches were previously reported before our
  observations ($\bigcirc$). Data for the pulse period and period
  derivative are from the ATNF Pulsar Catalogue
  (http://www.atnf.csiro.au/research/pulsar/psrcat/; version
  1.43).}\label{fig:ppdot_obs}
\end{center}
\end{figure}

Approximately 200 Galactic pulsars have been being observed regularly
from 1990 up until now by the 64-m radio telescope at the Parkes
Observatory located in western NSW,
Australia. Fig. \ref{fig:ppdot_obs} shows 165 pulsars that are in the
sample on the $P$ -- $\dot{P}$ diagram. Each pulse time-of-arrival
(TOA) for a particular pulsar was obtained every two to four
weeks. The data spans for these pulsars range from 5.3 to 20.8\,yr
with a total of $\sim$1911\,yr. By carrying out off-line data
reduction, 107 glitches were identified in 36 pulsars and 46 events
have not previously been published. Moreover, the recovery fraction
$Q$ and the timescale $\tau_{\rm d}$ were measured for 27 exponential
decays; post-glitch $\ddot\nu$ were also well measured for those
pulsars that exhibit long-term linear increase in $\dot\nu$ between
two adjacent glitches. More detailed results are being published in
\cite[Yu et al. (2012)]{ymh+}.

\section{Discussion}\label{sect:disn}

\begin{figure}
\begin{center}
\begin{tabular}{c}
\includegraphics[width=4.6cm,angle=-90]{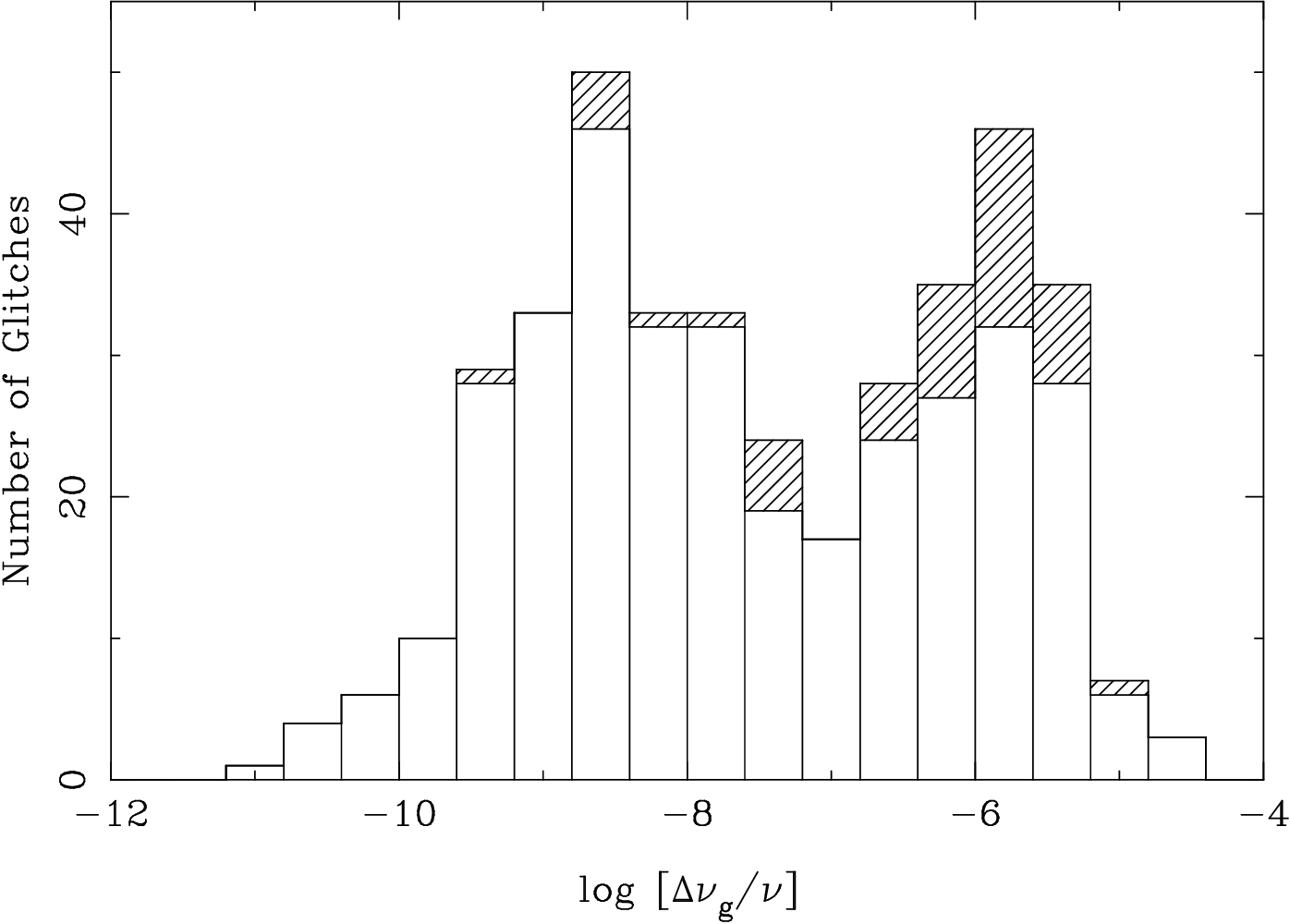} \\
\includegraphics[width=4.6cm,angle=-90]{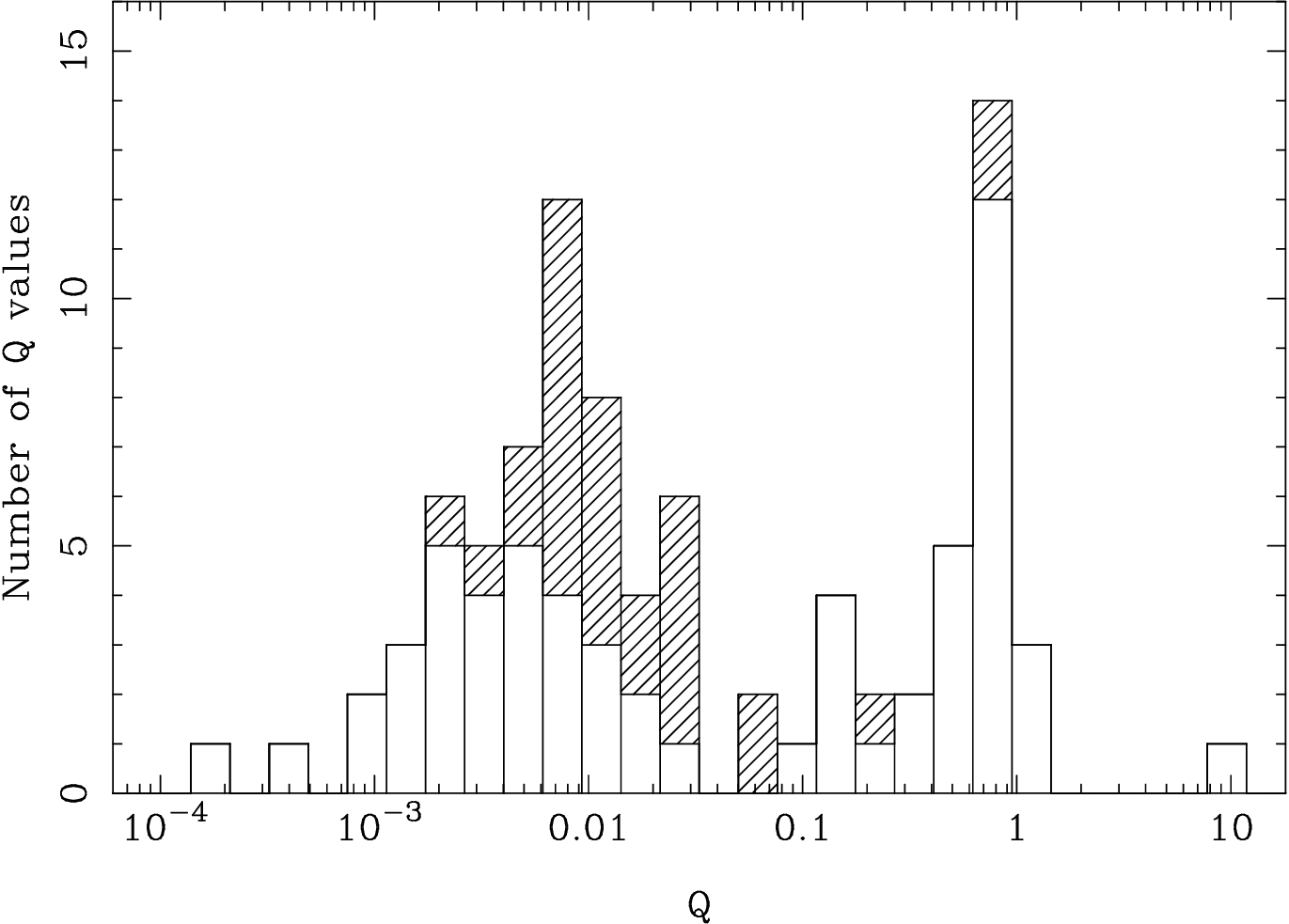} 
\end{tabular}
\end{center}
\caption{Histograms for the fractional glitch size $\Delta\nu_{\rm
    g}/\nu$ (upper panel) and the recovery fraction $Q$ of exponential
  decays (lower panel). In each plot, the data for the blank bars are
  from the literature, whereas those for the shaded bars are from Yu
  et al. (2012).}\label{fig:hist}
\end{figure}

Young pulsars have been thought to be appropriate samples to study
neutron-star interiors. It was soon after the discovery of the first
glitch that led to the realisation that neutron stars are
two-component rotators: a rigid-body bulk plus a faster-rotating
neutron superfluid \cite[(Baym et al. 1969)]{bppr69}. After
twenty-year-observation at Parkes, we now have further found i) the
bimodal distribution of the glitch fractional size $\Delta\nu_{\rm
  g}/\nu$ has been even clearer (upper panel, Fig. \ref{fig:hist}),
ii) glitches have frequently occurred in the pulsars that have
characteristic ages around 10\,kyr where, furthermore, large glitches
were generally seen, iii) exponential recoveries could be resolved
into multiple components corresponding to different timescales and iv)
post-glitch $\ddot\nu$ values are often positive, are generally larger
than the prediction from the magnetic-dipole model and are
proportional to the ratio between the slow-down rate and the
inter-glitch interval with a proportionality constant
$\sim$10$^{-3}$. These observations have further probed that neutron
stars may suffer the re-configuration of the crustal plate tectonics
and/or the substantial release of pinned vortices and, the rotational
equilibrium after a glitch established via vortex-drifting may occur
in multiple regimes resulting in the observed exponential and linear
recoveries.

Never-the-less, the available observations are still not sufficient to
entirely interpret the glitch phenomenon. As shown in the lower panel
of Fig. \ref{fig:hist}, the observed fraction $Q$ of the glitch that
recovers exponentially has exhibited a bimodal distribution with two
peaks locating at approximately 0.01 and 1.0 respectively. For the
two-component model, the conservation of angular momentum indicates
$Q=I_{\rm s}/I$, where $I$ is the total moment of inertia of the
neutron star and $I_{\rm s}$ the inertial moment of the superfluid
that gradually couples to the effective crust following a glitch. The
values of $Q \sim 1$ might have implied that, for some glitches,
$I_{\rm s}$ involved most of the neutron superfluid, though it has
been shown that the core superfluid may rigidly couple to the
realistic crust in a timescale $\lesssim$\,seconds \cite[(Alpar,
  Langer \& Sauls 1984; Alpar \& Sauls 1988)]{als84,as88}. Moreover, a
very large $Q \sim 8.7$ was detected in PSR~J1846$-$0258 recently
\cite[(Livingstone et al. 2010)]{lkg10} which is an even more
complicated situation. In addition, peculiar features have also been
observed. For instance, quasi-periodic oscillation in timing residuals
following a glitch has both been seen in PSR~B2334$+$61 \cite[(Yuan et
  al. 2010)]{ymw+10} and the Vela pulsar where, moreover, excess
delays were observed for the pulse arrival times measured at a lower
frequency \cite[(McCulloch et al. 1990)]{mhmk90}. Further exploration
on glitch-related phenomena may require intensive observations perhaps
with high precision and high observing cadence.

\section*{Acknowledgements}\label{sect:ack}

MY acknowledges the National Basic Research Program of China
(2012CB821800) and China Scholarship Council (No. 2009601129) for
funding. The Parkes radio telescope is part of the Australia
Telescope, which is funded by the Commonwealth of Australia for
operation as a National Facility managed by the Commonwealth
Scientific and Industrial Research Organisation (CSIRO).

\end{document}